\documentclass[9pt,final,letterpaper]{IEEEtran}
\IEEEoverridecommandlockouts

\usepackage{amsmath,amssymb,amsfonts}
\usepackage{mathdots}
\usepackage{mathtools}
\usepackage{commath}
\usepackage{microtype}

\usepackage{cite}
\newcommand{\myreferences}{extracted}

\usepackage[applemac]{inputenc}

\usepackage{tikz}
\usepackage{pgfplots}
\usetikzlibrary{shapes,fit,arrows,decorations.markings,matrix,plotmarks,positioning,fit}
\pgfplotsset{compat=newest}

\usepackage[font=footnotesize,caption=false]{subfig}

\usepackage{amsthm}
\newtheorem{theorem}{Theorem}
\newtheorem{corollary}{Corollary}
\newtheorem{lemma}{Lemma}
\newtheorem{proposition}{Proposition}
\newtheorem{remark}{Remark}
\newtheorem{assumption}{Assumption}

\newcommand{\ts}[1]{{\textnormal{#1}}}
\newcommand{\ie}{\emph{i.e.}\ }

\newcommand{\eg}{\emph{e.g.},}
\newcommand{\Rset}{\mathbb{R}}
\newcommand{\mb}{\mathbf}
\newcommand{\mc}{\mathcal}

\newcommand{\mbb}{\mathbb}

\title{Model predictive control of linear systems with preview information: feasibility, stability and inherent robustness}

\author{P. R. Baldivieso Monasterios and P. A. Trodden, \IEEEmembership{Member, IEEE}
\thanks{Work was supported by the Harry Nicholson PhD Scholarship, Department of Automatic Control \& Systems Engineering,
University of Sheffield.}
\thanks{P. R. Baldivieso Monasterios and P. A. Trodden are with the Department of Automatic Control \& Systems Engineering, University of Sheffield, Sheffield, UK {\tt\small \{prbaldivieso1, p.trodden\}@sheffield.ac.uk}}%
}

\begin{document}
\maketitle
\begin{abstract}
  The use of available disturbance predictions within a nominal model
  predictive control formulation is studied. The main challenge that
  arises is the loss of recursive feasibility and stability guarantees
  when a persistent disturbance is present in the model and on the
  system. We show how standard stabilizing terminal conditions may be
  modified to account for the use of disturbances in the prediction
  model. Robust stability and feasibility are established under the
  assumption that the disturbance change across sampling instances is
  limited.
\end{abstract}
      
\section{Introduction}

The stability and robustness of model predictive controllers has been
investigated extensively over the last few decades, and the
theoretical foundations are now mature~\cite{Mayne14}. Ultimately, the
purpose of any feedback controller is to reduce the effects of
uncertainty on a plant, and accordingly the MPC literature is---by and
large---separable into three, depending on whether or how uncertainty
is considered: nominal MPC for deterministic (\ie disturbance-free)
systems, robust MPC for systems subject to bounded uncertainty and
stochastic MPC for systems subject to uncertainties for which a
probabilistic description is available~\cite{KCbook}. While the
fundamental conditions for stabilizing nominal controllers have been
known for some time~\cite{MRR+00}, progress has been more recent in
robust and stochastic forms of MPC~\cite{Mayne14}. Even so, nominal
MPC remains the most popular choice in practical applications,
owing to its greater simplicity with respect to both design and
computation. The drawback to nominal MPC is, of course, that the
guarantees provided by the controller are indeed only nominal;
instances of zero robustness when the MPC controller employs a nominal
model but the real system is subject to an additive disturbance are
well documented~\cite{GMT+04}. Consequently, the inherent robustness
properties of nominal MPC have been investigated thoroughly, and
necessary and sufficient conditions are well
established~\cite{GMT+07,LAR+09}.

Despite these successes, a problem that has received relatively little
attention is that of regulation of a disturbed system by nominal MPC
when a \emph{forecast} or \emph{preview information} of future
disturbances is available for use in the prediction model. This
problem does not exactly fit the descriptions of robust or stochastic
MPC for regulation, but is closely related to topic of tracking,
wherein it has been known for some time that a disturbance model must
be included in the predictions in order to obtain offset-free
tracking~\cite{Muske2002}; in that context, different classes of
disturbance signals have been investigated, including piecewise
constant~\cite{CZ03,LAA+08,Muske2002} and
periodic~\cite{LAP+12,Pannocchia2005,Pannocchia2003}.

This \emph{preview control} problem~\cite{Sheridan1966} that we study
is motivated by applications in, \emph{inter alia}, power
systems~\cite{Warrington2018}, wind energy~\cite{LPS+11,KK13,LJR17},
water networks~\cite{Pereira2016} and automotive
control~\cite{GSW+14}. Another motivation is, as we show in
Section~\ref{sec:prelim}, the optimal control problem that arises in
\emph{distributed} MPC. The common feature is the availability of
(possibly accurate) predictions about an exogeneous disturbance acting
on the plant, obtained from sensors or otherwise: for example, wind
gusts acting on a turbine may be predicted by LIDAR, while the load
disturbance in a power network is partly deterministic (scheduled by
contract) and hence predictable. Most approaches to preview MPC focus
on formulating a controller in the setting of the application,
and then demonstrating the benefits to closed-loop performance of
including accurate preview information; stability guarantees are
largely lacking. An exception is the schemes based on robust MPC:
\cite{Pereira2016} proposes a controller that uses knowledge of a
deterministic demand in a tube-based robust MPC scheme. The approach
of \cite{Warrington2018} uses preview information to bound
uncertainties, and a parameterization of the control law in terms of
the disturbance is used obtain a robust controller.  The uncertain
load scheduling problem tackled by \cite{Neshastehriz2013} employs a
robust controller acting on the uncertain error dynamics, thus
eliminating the scheduled disturbance in the prediction model.

The aim of this note is to study the direct use of a disturbance
prediction within a \emph{nominal} MPC formulation for regulation, and
develop conditions for the recursive feasility, stability and inherent
robustness of such a controller without using the well-known
techniques from tracking or resorting to taking a robust or stochastic
approach. We consider a linear time-invariant, constrained system
subject to an additive disturbance, with the assumption that a
sequence of future disturbances (over the horizon of the controller)
is known to the controller at the current time, but---given the
non-deterministic nature of the disturbance---this sequence may change
arbitrarily (within a bounded set) at the next step, thus is
\emph{persistent}. It is known that nominal MPC controllers for linear
systems \emph{do} possess robustness to bounded additive
disturbances~\cite{LAR+09}, but these results are restricted to the
case where the prediction model omits the disturbance. Technical
challenges arise when the disturbance is included in the prediction
model, which mean that feasibility and stability are not assured, even
when standard stabilizing terminal conditions are employed.

Our contributions are the following: we modify standard terminal
ingredients for nominal MPC~\cite{RM_mpc_book} to account for the
disturbance appearing in the prediction model; we consider a general
form of terminal conditions permitting the use of a nonlinear terminal
control law. Under the modifications, the ill-posedness of the optimal
control problem is removed, and the desirable properties of invariance
of the terminal set and monotonic descent of the terminal cost hold
for the perturbed model. Then we study the inherent robustness (to
changing preview information) of the controlled system; we find that,
as in inherently robust nominal MPC, the system states converge to a
robust positively invariant set around the origin, the size of which
depends on the limit assumed (or imposed) on the step-to-step change
of the disturbance predictions. Illustrative examples show the
advantage of including the disturbance in the prediction model: the
region of attraction is larger than that for same controller omitting
the disturbance information.

The paper is organized as follows. Section~\ref{sec:prelim} defines
the problem and presents the basic MPC formulation that will be
studied. In Section~\ref{sec:terminal_cond}, modifications to the
optimal control problem, and the standard terminal conditions employed
in MPC, are developed to allow for the inclusion of the predicted
disturbance. In Section~\ref{sec:rec_feas}, feasibility and stability
of the system are studied, under both unchanging and changing
disturbances. Illustrative examples are given in
Section~\ref{sec:examples}, and conclusions are made in
Section~\ref{sec:conc}.

\emph{Notation and basic definitions}: $\mbb{R}_{+}$ is the set of
non-negative reals and $\mbb{N}$ is the set of positive integers. The
set of symmetric positive semidefinite $n \times n$ matrices is
$\mbb{S}^n_{+}$ and the set of positive definite ones is
$\mbb{S}^n_{++}$. We use normal face to denote a real-valued vector,
and boldface to denote a sequence of vectors. $|x|$ denotes a generic
norm of a vector $x \in \mbb{R}^n$; $\|x\|_Q^2$ denotes the quadratic
form $x^\top Q x$ in $x \in \mbb{R}^n$, where
$Q \in \mbb{R}^{n\times n}$.  For sets $\mc{A} \subset \mbb{R}^n$ and
$\mc{B} \subset \mbb{R}^n$, the Minkowski sum is
$\mc{A} \oplus \mc{B} \triangleq \{ a + b: a \in \mc{A}, b \in
\mc{B}\}$. The notation $A^{-1}(X)$ represents the preimage of the set
$X$ under the matrix $A$,
\ie~$A^{-1}(X) \triangleq \{ x : Ax \in X \}$. A set
$X \subseteq \mbb{R}^n$ is positively invariant for a (autonomous)
system $x^+ = f(x)$, where $f\colon \mbb{R}^n \mapsto \mbb{R}^n$, if
$x \in X$ implies $f(x) \in X$. A set $X \subseteq \mbb{R}^n$ is
control invariant for a (non-autonomous) system $x^+ = f(x,u)$ and
input set $\mbb{U}$, where
$f\colon \mbb{R}^n \times \mbb{R}^m \mapsto \mbb{R}^n$, if for all
$x \in X$ there exists a $u \in \mbb{U}$ such that $f(x,u) \in X$. A
C-set is a closed and bounded (compact) convex set containing the
origin; if the origin is in the interior, then it is a PC-set. An LTI
system $x^+ = Ax + Bu$ is reachable (controllable from the origin) if,
for any $x_1 \in \mbb{R}^n$, there exists a sequence of controls that
steers the state from $x(0) = 0$ to $x_1$ in finite time. When it is
clear from the context, and in order to keep notation simple, we
reserve $k$ as the sampling instant and $i$ as the prediction step;
that is, $x(k)$ denotes a state at sampling instant $k$, while $x(i)$
denotes an $i$-step ahead prediction of $x$; where confusion may
arise, we write $x(k+i)$ to denote the latter.
\section{Problem setup and basic formulation}
\label{sec:prelim}
\subsection{System description and control objective}

Consider the regulation problem for a discrete-time, linear time-invariant
(LTI) system subject to an additive disturbance:
\begin{equation}
	x^+ = Ax + Bu + w
        	\label{eq:lss}
\end{equation}
where $x\in\Rset^n$, $u\in\Rset^m$, $w\in\Rset^n$ are, respectively,
the state, input, and disturbance at the current time, and $x^+$ is
the successor state. The control objective is to regulate the system
state $x \in \Rset^n$ to the origin, despite the action of the
disturbance, while satisfying any constraints on states and
inputs:
\begin{equation*}
	x\in\mbb{X} \subset \mbb{R}^n, \quad u\in\mbb{U} \subset \mbb{R}^m
\end{equation*}
\begin{assumption}[Basic system assumptions]\label{assump:contr}
  The pair $(A,B)$ is reachable, and the matrices $(A,B)$ are
  known.
\end{assumption}
\begin{assumption}[Constraint sets]\label{assump:constraint_sets}
  The sets $\mathbb{X}$ and $\mathbb{U}$ are PC-sets; the set
  $\mbb{W}$ is a C-set.
\end{assumption}
We make the following assumption about the preview information available to
the controller at the time it makes control decisions:

\begin{assumption}[Information available to the controller]\label{assump:dist}~
\begin{enumerate}
  \item The state $x(k)$ and disturbance $w(k)$ are known exactly at
  time $k$; the future disturbances are not known exactly but satisfy
  $w(k+i) \in \mbb{W}, i \in \mbb{N}$.
  \item  At any time step $k$, a \emph{prediction} of future disturbances,
    over a finite horizon, is available.
    \end{enumerate}
\end{assumption}
Full state availability is a standard assumption, but disturbance
measurement availability is not; this might be restrictive but can
arise in problems in, for example, power systems and, as we show in
Section~\ref{sec:dmpc}, distributed MPC.

The main focus of the note is on the second part of
Assumption~\ref{assump:dist}: the availability of disturbance
predictions. Our aim is to study how such information can be used
within a nominal MPC formulation, and establish conditions for
feasibility, stability and inherent robustness of the resulting
controller and closed-loop system. Note that that we do not assume
anything about the \emph{accuracy} of the disturbance predictions, and
in fact will permit the predictions to vary over time (in which case
it is implicit that earlier predictions were not accurate).
\subsection{A basic MPC formulation with disturbance information}
\label{sec:mpc}
We study an MPC formulation that would arise from the direct (and
perhaps na\"ive) inclusion of disturbance predictions in a nominal MPC problem for regulation, without using techniques from
tracking MPC to handle it. As per the preceding assumptions, the
controller has knowledge of the current state $x$ and disturbance $w$
but, additionally, a \emph{prediction} of the future disturbances. We
define the disturbance prediction at the current time to be a sequence
$\mb{w} \triangleq \left\{w(0),w(1),\ldots,w(N-1)\right\}$, where
$w(0) = w$ (the measured disturbance), with that---and each future
value of the sequence---lying within the bounded set $\mbb{W}$. The
basic optimal control problem for $(x,\mb{w})$ is
\begin{equation*}
	\mbb{P}(x;\mb{w}) \colon \min\bigl\{V_N(x,\mb{u};\mb{w}):\mb{u}\in\mc{U}_N(x;\mb{w})\bigr\}
\end{equation*}
where $\mc{U}_N(x;\mb{w})$ is defined by, for $i = 0\dots N-1$,
	\begin{align*}
		x(0) &= x, \\
		x(i+1) &= Ax(i) + Bu(i) +w(i),\\
		x(i) &\in \mbb{X},\\
  		u(i) &\in \mbb{U},\\
         x(N) & \in \mbb{X}_f.
	\end{align*}
%
In this problem, the cost function comprises, in the usual way, a
stage cost plus terminal penalty:
\begin{equation*}
  V_N(x,\mb{u};\mb{w}) \triangleq V_f\bigl(x(N);\mb{w}\bigr) + \sum_{i = 0}^{N-1} \ell\bigl( x(i), u(i); \mb{w} \bigr),
\end{equation*}
where $\ell\colon \mbb{R}^n \times \mbb{R}^m \mapsto \mbb{R}_{+}$,
$V_f\colon \mbb{R}^n \mapsto \mbb{R}_{+}$; both depend, in a way that
is to be defined, on the disturbance sequence. The domain of this
problem
$\mc{X}_N(\mb{w}) \triangleq \left\{ x \in \mbb{X}:
  \mc{U}_N(x;\mathbf{w}) \neq \emptyset \right\}$, and depends on
$\mb{w}$. The objective of the controller is to steer the system
towards its equilibrium despite the disturbances; when the disturbance
is \emph{persistent} then regulation is merely to a neighbourhood of
the origin. In this case, we aim to achieve regulation to a
neighbourhood of the origin using only nominal MPC, \ie~avoiding
a robust formulations.

The solution of this problem at a state $x$ and with disturbance
sequence $\mb{w}$ yields an optimal control sequence
\begin{equation*}
	\mb{u}^0(x;\mb{w}) \triangleq \bigl\{ u^0(0;x,\mb{w}), u^0(1;x,\mb{w}), \dots, u^0(N-1;x,\mb{w}) \bigr\}.
\end{equation*}
The application of the first control in the sequence defines the
control law $u = \kappa_N(x;\mb{w}) = u^0(0;x,\mb{w})$. At the
successor state $x^+ = Ax + B\kappa_N(x;\mb{w}) + w(0;\mb{w})$, the
problem is solved again, yielding a new control sequence. Motivated by
the limited accuracy of any disturbance prediction, the disturbance
sequence used in the problem at $x^+$ may change arbitrarily to
$\mb{w}^+$, and is not necessarily equal to the tail of the previous
predictions $\mb{w}$.

\subsection{A motivating application: distributed MPC}
\label{sec:dmpc}

    Our problem description can arise naturally in distributed
    forms of MPC. Consider the two dynamically coupled subsystems
    \begin{align*}
      x_1^+ = A_{11} x_1 + B_1 u_1 + A_{12} x_2\\
      x_2^+ = A_{22} x_2 + B_2 u_2 + A_{21} x_1
      \end{align*}
      and suppose each subsystem controller $p\in \{1,2\}$ has an
      optimal control problem of the form
      \begin{equation*}
	\mbb{P}_p(x_p;\mb{w}_p) \colon \min\bigl\{V_{p,N}(x_p,\mb{u}_p;\mb{w}_p):\mb{u}_p\in\mc{U}_{p,N}(x_p;\mb{w}_p)\bigr\}
\end{equation*}
where $\mc{U}_{p,N}(x_p;\mb{w}_p)$ is defined by 
	\begin{align*}
		x_p(0) &= x_p, \\
		x_p(i+1) &= A_{pp}x_p(i) + B_pu_p(i) + w_p(i),\\ 
		x_p(i) &\in \mbb{X}_p,\\ 
  		u_p(i) &\in \mbb{U}_p,\\
         x_p(N) & \in \mbb{X}_{p,f}.
	\end{align*}%
Suppose that, at some sampling instant, controller $1$ solves its
problem first (assuming some information about $\mb{w}_1$---in the
simplest case $\mb{w}_1 = \mb{0}$) to obtain an optimized future
control sequence $\mb{u}^0_1(x_1)$ and associated sequence of state
predictions, $\mb{x}^0_1(x_1)$. If the controller then shares these
predictions with controller $2$, then controller $2$ receives preview
information $\mb{w}_2 = \mb{x}^0_1(x_1)$ available to use in its own
problem. In fact, $\mb{w}_2$ fits exactly with
Assumption~\ref{assump:dist}, for the first element $w_2(0) = x_1$ is
known accurately, while the remaining elements are subject to change
at the next sampling instant (after controller $1$ solves its
optimization problem once more).

Distributed MPC schemes essentially differ in how they use this
information: it is known that the direct inclusion into the MPC
formulation, as we have illustrated here, is problematic, and therefore
iterative~\cite{VHR+08} and robust~\cite{FS12,TM17} methods are
typically used.

\subsection{Stabilizing conditions for nominal MPC and issues that arise from including preview information}
When the prediction model omits the disturbance, an established
approach to ensuring desirable properties of the controller and
closed-loop system is to design the cost and terminal set to satisfy
some standard conditions~\cite{RM_mpc_book}. In particular, when the
model employed is $x^+ = Ax + Bu$ and neither the terminal set nor the
cost function depend on $\mb{w}$, the latter are constructed to
satisfy the following:
\begin{assumption}[Terminal set properties]\label{assump:Xf}
The set $\mbb{X}_f \subset \mbb{X}$ is a PC-set.
  \end{assumption}

\begin{assumption}[Cost function bounds]\label{assump:costs}
  The functions $\ell(\cdot,\cdot)$, $V_f(\cdot)$ are continuous, with $\ell(0,0) = 0$, $V_f(0) = 0$ and such that, for some $c_1 > 0$, $c_2 > 0$, $a > 0$,
  \begin{align*}
    \ell(x,u) &\geq c_1 | x |^a \ \text{for all} \ x \in \mc{X}_N, u \in \mbb{U}\\
    V_f(x) &\geq c_2 |x|^a \ \text{for all} \ x \in \mbb{X}_f          
    \end{align*}
  \end{assumption}
\begin{assumption}[Basic stability assumption]
  For all $x \in \mbb{X}_f$,
  \begin{equation*}
    \min_{u \in \mbb{U}} \bigl\{ V_f(Ax+Bu) + \ell(x,u) : Ax+Bu \in \mbb{X}_f\bigr\} \leq V_f(x).
    \end{equation*}
\label{assump:basic}
\end{assumption}
\begin{assumption}[Control invariance of
  $\mbb{X}_f$]\label{assump:invariance}
  The set $\mbb{X}_f$ is control invariant for $x^+ = Ax+Bu$ and the
  set $\mbb{U}$.
\end{assumption}
In turn, each of the \emph{controllability sets}
$\mc{X}_0, \mc{X}_1, \dots, \mc{X}_N$, where
\begin{equation*}
\mc{X}_{i+1} = \{ x \in \mbb{X} : \exists u \in \mbb{U} \ \text{such that} \ Ax + Bu \in \mc{X}_i \} ,
  \end{equation*}
  with $\mc{X}_0 = \mc{X}_f$, is control invariant for $x^+ = Ax + Bu$ and $\mbb{U}$, and these sets are \emph{nested}: $\mc{X}_N \supseteq \mc{X}_{N-1} \supseteq \dots \supseteq \mc{X}_1 \supseteq \mc{X}_0$. Furthermore, the set $\mc{X}_N$ (and $\mc{X}_{N-1}$) is
  positively invariant for $x^+ = Ax + Bu$ under the MPC control law
  $u = \kappa_N (x)$ and admissible with respect to the
  constraints. Consequently, if $x(0) \in \mc{X}_N$ then (i) the
  optimal control problem is recursively feasible, (ii) constraints
  are satisfied for all times, and the origin is asymptotically stable
  for the closed-loop system, with region of attraction $\mc{X}_N$~\cite{RM_mpc_book}. 

When disturbances are present in the model, the same  ingredients fail to attain the same guarantees, and several issues arise. 
    \begin{itemize}
    \item \emph{Ill-posed objective}. For the perturbed dynamics,
      $(x=0,u=0)$ is not, in general, an equilibrium pair for the
      system. Consequently, employing a cost function that satisfies
      Assumptions~\ref{assump:costs} and~\ref{assump:basic} will
      \emph{not} ensure that the value function satisfies the
      conditions of a Lyapunov function.
    \item \emph{Loss of invariance of $\mbb{X}_f$}. Control invariance
      of $\mbb{X}_f$ for $x^+ = Ax+Bu$ does not imply control
      invariance for $x^+ = Ax+Bu+ w$, even when $w$ is
      constant. Invariance of $\mbb{X}_f$ is and is a \emph{necessary}
      condition for the controllability sets to be control invariant
      and nested~\cite{Mayne_apol} when state constraints $\mbb{X}$
      are considered.
    \item \emph{Loss of nesting of the controllability sets}. The
      controllability sets are defined by the iteration
      \begin{equation}
        \mc{X}_{i+1} = \mbb{X} \cap A^{-1}\bigl( \mc{X}_i \oplus -B\mbb{U} \oplus - \{w(N-i)\} \bigr),
		\label{eq:nesty}
              \end{equation}
              with $\mc{X}_0 = \mbb{X}_f$. The final summand
              in~\eqref{eq:nesty} is problematic, for $\{w(N-i)\}$ is
              a point, not a set containing the origin, and therefore
              induces \emph{translation} of the controllability sets
              between iterations. Moreover, if $A$ is unstable, then
              with each iteration the controllability set is shifted
              further away as disturbance propagates through the
              unstable predictions. The implication is that $\mc{X}_N$
              is not control invariant for $x^+ = Ax + Bu + w$, and
              recursive feasibility may not be easy to establish, even
              if $\mc{X}_f$ is control invariant for the terminal
              dynamics.
  \end{itemize}

\section{Modifications to the cost and terminal ingredients}
\label{sec:terminal_cond}
In this section, we develop modifications to the cost
$\ell(\cdot,\cdot)$ and terminal ingredients $V_f(\cdot)$, $\mbb{X}_f$
for the model predictive controller that aim to overcome the first two
of the fundamental issues outlined in the previous section. With
respect to the third issue, we show in Section~\ref{sec:rec_feas} that
recursive feasibility of the control problem can be established
without relying on nestedness of the controllability sets; in fact, we
show that a different kind of nested property holds here. 

Our approach is to modify the cost and terminal set using the
available preview information. We assume a constant-disturbance
terminal prediction model, take stabilizing ingredients designed for
its nominal counterpart, and translate the costs and terminal set to
account for the non-zero equilibrium caused by the disturbance; as we
shall show, the resulting ingredients achieve the required properties
of invariance and monotonicity with respect this new predicted
equilibrium instead of the origin. The results
in this section are essentially derived from known results on
stabilizing MPC~\cite{RM_mpc_book}, but lay the foundations for our
  main results in Section~\ref{sec:rec_feas}.

\subsection{Constant-disturbance terminal dynamics}
In order to synthesize appropriate terminal ingredients, one first
needs to consider the terminal (beyond the horizon) dynamics of the
prediction model. This motivates the next assumption, regarding the
disturbance sequence: by definition, the disturbance predictions are
known only up to $N$ steps ahead in the future, and therefore to
consider the terminal dynamics we setup the predicted disturbance
sequence in the following way:
\begin{assumption}\label{assump:wf}
  An \emph{admissible} disturbance sequence
  $\mb{w} = \{w(0), \allowbreak \dots ,w(N-1)\}$ has
  $w(i) \in \mbb{W}$, $i=0\dots (N-2)$, and
  $w(N-1) \in \mbb{W}_f \subseteq \mbb{W}$; we write
  $\mb{w} \in \mc{W} \triangleq \mbb{W} \times \dots \times \mbb{W}
  \times \mbb{W}_f$.  To account for $i \geq N$, the sequence is continued indefinitely by concatenation with $\mb{w}_f \triangleq \bigl\{ w(i) \bigr\}_{i \geq N}$, with  $w(i) = w_f(\mb{w}) \triangleq w(N-1)$ for $i \geq N$.
\end{assumption}
That is, the predicted disturbance is, after $N$ steps, constant and
equal to $w_f(\mb{w}) = w(N-1)$, the final value in the $N$-length
sequence $\mb{w}$. This final value lies in a subset
$\mbb{W}_f \subseteq \mbb{W}$; thus,
$w_f\colon \mc{W} \mapsto \mbb{W}_f$. A constant terminal disturbance
assumption is not necessarily restrictive; for example,
\cite{Pannocchia2005} considers disturbance models that are
integrating and reach a constant value.

Assumption~\ref{assump:wf} implies that, at a time instant $k$ and
given the disturbance prediction $\mb{w}$, the terminal prediction
dynamics are given by $x^+ = Ax + Bu + w_{f}(\mb{w})$, with
$w_{f}(\mb{w})$ constant for all prediction steps $i \geq N$. At the
next time instant, however, $w_f(\mb{w}^+)$ need not be equal to
$w_f(\mb{w})$, and we address this issue in
Section~\ref{sec:rec_feas}. Until then, for simplicity of notation we
refer to $w_f(\mb{w})$ as $w_f$.
\subsection{Equilibria of the terminal dynamics}
To proceed, we first assume the availability of terminal ingredients
for the nominal system; in the sequel, the overbar notation denotes a
function or set pertaining to the nominal dynamics $x^+ = Ax + Bu$.
\begin{assumption}
  There is known a set $\bar{\mbb{X}}_f$ and functions
  $\bar{\ell}(\cdot,\cdot)$, $\bar{V}_f(\cdot)$ satisfying
  Assumptions~\ref{assump:Xf}--\ref{assump:invariance} for the
  dynamics $x^+ = Ax+Bu$ and constraint sets
  $(\beta_x \mbb{X}, \beta_u \mbb{U})$, where
  $\beta_x, \beta_u \in [0,1)$.
\label{assump:terminal_law}
\end{assumption}
The set $\bar{\mbb{X}}_f$ is a control invariant set (for the nominal dynamics) that resides within $\beta_x \mbb{X}$, which is the state constraint set scaled to allow for the effect of non-zero disturbances on the terminal dynamics; the selection of suitable scaling factors $\beta_x$ and $\beta_u$ is described in the sequel. Assumption~\ref{assump:terminal_law} implies the existence of a (possibly set-valued) control law, $u_f\colon \mbb{R}^n \mapsto 2^{\mbb{R}^m}$ that induces invariance of the set $\bar{\mbb{X}}_f$ with respect to the nominal dynamics.
\begin{multline}
  u_f(x) = \bigl\{u\in\beta_u \mbb{U} :  Ax+Bu \in\bar{\mbb{X}}_f,\\
  \bar{V}_f(Ax+Bu)+\bar{\ell}(x,u)\leq \bar{V}_f(x)\bigr\}, \
  \text{for} \ x \in \bar{\mbb{X}}_f.
\label{eq:terminal_law}
\end{multline}
We denote the system $x^+ = Ax+Bu$ under the control law $u = \bar{\kappa}_f(x)$---where, if necessary, $\bar{\kappa}_f(x)$ is an appropriate selection from $u_f(x)$, but is otherwise equal to $u_f(x)$---as $x^+ = f_{\bar{\kappa}_f}(x) = Ax + B\bar{\kappa}_f(x)$ (so that $f_{\bar{\kappa}_f}\colon \bar{\mbb{X}}_f \mapsto \bar{\mbb{X}}_f$). The following result is an immediate consequence of the existence of $\bar{\kappa}_f(\cdot)$ satisfying the conditions of Assumption~\ref{assump:Xf}--\ref{assump:invariance}.
\begin{lemma}[Stabilizing terminal control law]\label{lem:kfstab}
  Suppose Assumptions~\ref{assump:contr} and~\ref{assump:terminal_law}
  hold. The origin is an asymptotically stable equilibrium for the
  system $x^+ = f_{\bar{\kappa}_f}(x)$, with region of attraction $\bar{\mbb{X}}_f$.
\end{lemma}
The difficulty with translating the terminal ingredients to the new
steady state or equilibrium point of the system perturbed by $w_f$ is
that this point is not immediately known, since the terminal control
law is nonlinear and/or set-valued. The approach we take is to
linearize the terminal dynamics and control law around the equilibrium
point $x=0 \in \bar{\mbb{X}}_f$, for which we make the following
assumption.
\begin{assumption}\label{assump:diff}
  $\bar{\kappa}_f(\cdot)$ is continuous
over $\bar{\mbb{X}}_f$, with $\bar{\kappa}_f(0) = 0$, and continuously
  differentiable in a neighbourhood of $x=0$.
\end{assumption}
In reality, this is a mild assumption: a ready choice of $\bar{\kappa}_f(\cdot)$ is a linear, stabilizing $K$. Even if $\bar{\kappa}_f(\cdot)$ is piecewise linear, as in~\cite{GP13}, in order to maximize the size of $\bar{\mbb{X}}_f$, the control law is typically linear (for a linear system) in a neighbourhood of the origin when $\mbb{X}$ and $\mbb{U}$ are PC-sets. Then the linearization of $x^+ = f_{\bar{\kappa}_f}(x)$ yields
\begin{equation}
\Pi \triangleq \left.\dpd{\bar{\kappa}_f(x)}{x}\right|_{x=0};\quad \Phi \triangleq \left.\dpd{\bigl( Ax + B\bar{\kappa}_f(x) \bigr)}{x}\right|_{x=0} = A + B\Pi
\label{eq:Phi_Pi} 	
\end{equation}
from which the equilibrium point of the linearized system
$x^+ = \Phi x +  w_f$ is
\begin{align*}
  x_f(w_f) = \Psi  w_f, &&  u_f(w_f) = \Pi\Psi  w_f
\end{align*}
where $\Psi \triangleq \bigl(I - \Phi \bigr)^{-1}$. In view of Lemma~\ref{lem:kfstab} and Assumption~\ref{assump:diff}, $\Phi = (A+B\Pi)$ is well defined and strictly stable, hence $\Psi = (I-\Phi)^{-1}$ exists and is unique. This proves the following.
\begin{lemma}[Existence of equilibrium]\label{lem:xeue}
  Suppose Assumptions~\ref{assump:contr}, \ref{assump:terminal_law}
  and \ref{assump:diff} hold. The point
  $x_{f}(w_f)$ exists and is the unique equilibrium of $x^+ = \Phi x +  w_f$, with $w_f$ constant.
  \end{lemma}
\subsection{Modified terminal conditions: translation and properties}
We propose to translate the cost function, terminal sets and control law to this new point $(x_f,u_f)$:
\begin{align*}
      \ell(x,u;w_f) &\triangleq \bar{\ell}\bigl( x - x_f(w_f), u - u_f(w_f) \bigr)\\
      V_f(x;w_f) &\triangleq \bar{V}_f\bigl(x - x_f(w_f)\bigr)\\
      \mbb{X}_f(w_f) &\triangleq \bar{\mbb{X}}_f \oplus \bigl\{x_f(w_f)\bigr\}\\
      \kappa_f(x;w_f) &\triangleq \bar{\kappa}_f\bigl(x - x_f(w_f)\bigr) + u_f(w_f)
      \end{align*}
      The translated sets and functions are then the ones employed in
      the optimal control problem with perturbed prediction model. The
      main result of this section then establishes that the resulting
      terminal ingredients satisfy the required conditions
      (counterparts to Assumptions \ref{assump:Xf}--\ref{assump:basic}) for
      $x^+ = Ax + Bu +  w_f$. First, the following assumption is
      required.
\begin{assumption}\label{assump:alpha}
  There exist scalars
  $\alpha_x, \alpha_u \in [0,1)$ such that
  \begin{align*}
    \Psi \mbb{W}_{f}\subseteq\alpha_x\mbb{X}&&
    \Pi \Psi \mbb{W}_{f} \subseteq\alpha_u\mbb{U}
  \end{align*}
\end{assumption}
This assumption essentially limits the size of the terminal disturbance
set, $\mbb{W}_f$, with respect to the state and input constraint
sets. Then the following result restores the stability properties.
\begin{proposition}
  Suppose that Assumptions~\ref{assump:contr}--\ref{assump:dist},
  \ref{assump:wf}--\ref{assump:alpha} hold. For any fixed
  $w_{f}\in \mbb{W}_f$, (i) the set
\begin{equation}
  \mbb{X}_f(w_{f}) = \bar{\mbb{X}}_f\oplus\bigl\{ x_f(w_f) \bigr\}
\label{eq:terminal_set_affine}
\end{equation} 
is positively invariant for $x^+ = Ax+B\kappa_f(x;w_f) + w_{f}$, and (ii) the functions $\ell(x,u;w_f)$ and $V_f(x;w_f)$ satisfy
\begin{multline*}
V_f\bigl([Ax+B\kappa_f(x;w_f)+w_f]; w_f\bigr) + \ell\bigl(x,\kappa_f(x;w_f); w_f\bigr) \\ \leq V_f(x;w_f) \ \text{for all} \ x \in \mbb{X}_f(w_f).
  \end{multline*}
  Furthermore,  (iii) $\mbb{X}_f(w_{f}) \subseteq \mbb{X}$, and $\kappa_f(\mbb{X}_f;w_f) \subseteq \mbb{U}$ if $\alpha_x + \beta_x\leq1$ and $\alpha_u + \beta_u\leq1$.
  Finally, (iv) the point $x_f(w_f)$ is an
  asymptotically stable equilibrium point for the system
  $x^+ = Ax + B\kappa_f(x;w_f) + w_f$. The region of attraction is
  $\mbb{X}_f(w_f)$.
\label{thm:terminal_set_inv}
\end{proposition}

This result implies that the set $\mbb{X}_f(w_f)$ is control invariant
for the system $x^+ = Ax + Bu+w_f$, $w_f$ constant, and input set
$\mbb{U}$. In particular, the set is positively invariant for
$x^+ = Ax+B\kappa_f(x;w_f)+w_f$. Within this set, $V_f(\cdot)$ is a
control Lyapunov function. Thus, we have established stability of the
predicted terminal dynamics to an equilibrium point under the
assumption of a constant disturbance $w_f$. In the next section, we
consider the receding-horizon implementation of the optimal control
problem, allowing the terminal disturbance to change across sample
times but, at each sampling instant, remain constant beyond the end of
the horizon.

\begin{remark}[Offset-free regulation and links with tracking and
  robust MPC]
  In view of the regulation objective, the stabilization of the
  perturbed system under constant disturbances to
  $x_f(w_f) = \Psi w_f$ may seem unambitious. Ultimately, however,
  this is an issue of available degrees of freedom: if $x = 0$ is
  required to be an equilibrium of $x^+ = Ax + Bu + w_f$, then
  standard results from offset-free tracking inform us this is
  possible, in general, if and only if $m \geq n$---a strong condition
  that is typically not met unless the system is over-actuated. A
  pragmatic solution, then, can be to define some outputs
  $y = Cx \in \mbb{R}^p$, with $p \leq m$, and steer the states to
  values that satisfy $y_f = Cx_f = 0$ (or minimize $||Cx_f||^2$). With
  the regulation goal in mind, such a scheme requires careful (and
  application-dependent) design of $C$ in order to ensure that some
  states do not converge to values far from the origin.
  
  When such an option is not pursued, there is an interesting link
  with what is known from robust MPC. The smallest neighbourhood
  of the origin that the states of a perturbed system
  $x^+ = \Phi x + w$, under bounded $w \in \mbb{W}$, can stay within
  is the minimal robust positively invariant (mRPI) set. given by the Minkowski summation
  \begin{equation*}
    \mc{R}_\infty \triangleq \bigoplus_{i = 0}^{\infty} \Phi^i  \mbb{W}.
  \end{equation*}
  There is a strong connection between this set and the equilibrium
  point $x_f(w_f) = \Psi w_f$ under a constant disturbance $w_f$:
  $x_f(w_f) \in \mc{R}_{\infty}$ and, similar to this summation,
  \begin{equation*}
      x_f(w_f) = (I-\Phi)^{-1} w_f = (I + \Phi + \Phi^2 + \dots)  w_f = \bigoplus_{i = 0}^{\infty} \Phi^i  \{w_f\}.
    \end{equation*}
  \end{remark}

\section{Feasibility, stability and inherent robustness under changing disturbances}
\label{sec:rec_feas}
In this section, we analyse the properties of the MPC controller with
the proposed cost and terminal set modifications, considering the
possibility that the disturbance sequence provided to the controller
changes over time. Given a disturbance sequence $\mb{w}$, the set of
states for which the problem $\mbb{P}(x;\mb{w})$ is feasible is
$\mc{X}_N(\mb{w}) =\{x\in\mbb{X}:
\mc{U}_N(x;\mb{w})\neq\emptyset\}$. The challenge that arises is a
consequence of the receding horizon implementation of MPC: if
$\mbb{P}(x;\mb{w})$ is feasible and yields a solution
$\mb{u}^0(x;\mb{w}) \triangleq \{u^0(0;x,\mb{w}), u^0(1;x,\mb{w}),
\dots, u^0(N-1;x,\mb{w})\}$, then the first control in this sequence
is applied (the implicit control law is
$\kappa_N(x;\mb{w}) \triangleq u^0(0;x,\mb{w})$), the system evolves
to $x^+ = Ax + B\kappa_N(x;\mb{w}) + w$, and the problem to be solved
at the subsequent time is $\mbb{P}(x^+;\mb{w}^+)$. The question is,
when is this problem feasible, given that the disturbance sequence may
have changed (perhaps arbitrarily) from $\mb{w}$ to $\mb{w}^+$?

\subsection{Unchanging disturbance sequences}
Our first results on feasibility and stability consider the artificial
situation where the disturbance sequence is shifted in time but
otherwise unchanged, \ie the sequence $\mb{w}^+$ is the \emph{tail} of
the sequence $\mb{w}$, and the disturbance acting on the plant is
equal to the first element in the sequence at each sampling
instant. These results, though immediate consequences of the
preceding developments and standard results in MPC, form
the basis for the more realistic case of a persistently changing
disturbance sequence considered in Sec.~\ref{sec:main2}.

In order to present the results, we need to define some notation. For
a disturbance sequence
$\mb{w} = \bigl\{ w(0), w(1), \dots,w(N-2), w(N-1) \bigr\}$, the
associated tail is
$\tilde{\mb{w}}(\mb{w}) \triangleq \left\{
  w(1),\dots,w(N-1),w(N-1)\right\}$ where the first $N-1$ elements of
$\mb{w}$ are augmented by continuing the final value $w(N-1)$ for one
further step. The infinite-length disturbance sequence $\mb{w}_N(k)$
is formed from concatenating $\mb{w}(k)$ and $\mb{w}_f(k)$ where
  \begin{equation*}
  \mb{w}_N(k) = \bigl\{ w\bigl(0;\mb{w}(k)\bigr), \dots, w\bigl(N-1;\mb{w}(k)\bigr), \mb{w}_f(k) \bigr\}.
\end{equation*}
  The latter is the infinite sequence of constant
  disturbances obtained by holding the final value of
  $\mb{w}(k)$---that is,
  $\mb{w}_f(k) = \bigl\{ w_f\bigl( \mb{w}(k) \bigr) \bigr\}$ where
  $w_f\bigl( \mb{w}(k) \bigr) = w\bigl(N-1;\mb{w}(k) \bigr)$. Finally, $\mb{w}_i(k)$, for $i = 0 \dots N$, is a version of $\mb{w}_N(k)$ of omitting the first $N-i$.

The next result, and those in the remainder of the paper, suppose
  that Assumptions~\ref{assump:contr}--\ref{assump:dist},
  \ref{assump:terminal_law}--\ref{assump:alpha} hold.

\begin{proposition}[Feasibility under unchanging disturbance]\label{prop:rec_feas}   
  Let $\mb{w} = \{w(0),w(1),\allowbreak \dots,\allowbreak w(N-1)\} \in \mc{W}$. If $\mb{w}^+ = \tilde{\mb{w}}(\mb{w})$, then (i) $x \in \mc{X}_N(\mb{w})$ implies
    $x^+ = Ax + B\kappa_N(x;\mb{w}) + w(0;\mb{w}) \in \mc{X}_N(\mb{w}^+)$.
(ii) Given $\mb{w}(0)$, the set
  \begin{equation*}
    \bigcup_{k=0}^{\infty} \mc{X}_N\big(\mb{w}(k)\bigr) \ \text{with} \ \mb{w}(k+1) = \tilde{\mb{w}}\bigl(\mb{w}(k)\bigr) 
  \end{equation*}
  is control invariant for $x(k+1) = Ax(k) + Bu(k) + w(k)$ and $\mbb{U}$, where $w(k) = w\bigl(0;\mb{w}(k)\bigr)$. (iii) The unions of controllability sets are nested:
  \begin{equation*}
    \bigcup_{k=0}^{\infty} \mc{X}_N\bigl(\mb{w}_N(k)\bigr) \supseteq \bigcup_{k=0}^{\infty} \mc{X}_{N-1}\bigl(\mb{w}_{N-1}(k)\bigr) \supseteq \dots \supseteq \bigcup_{k=0}^{\infty} \mc{X}_0\bigl(\mb{w}_0(k)\bigr),
  \end{equation*}
  where 
  \begin{equation*}
	\mc{X}_{i+1}(\mb{w}_{i+1}) = \mbb{X} \cap A^{-1}\bigl( \mc{X}_i(\mb{w}_i) \oplus -B\mbb{U} \oplus - \{w(N-i)\} \bigr),
\label{eq:feasible_region}
\end{equation*}
with $\mc{X}^0(\mb{w}_0) = \mbb{X}_f(\mb{w}_f)$, and $\mb{w}_i(k+1) = \tilde{\mb{w}}\bigl(\mb{w}_i(k)\bigr)$.
\end{proposition}

\begin{remark}
  Whereas the individual sets
  $\mc{X}_N, \mc{X}_{N-1}, \dots, \mc{X}_0$ are not necessarily
  control invariant and nested, their unions are, provided the
  disturbance sequence updates by taking the tail.
\end{remark}

\begin{remark}
  Owing to the nilpotency of the dynamics of the disturbance sequence (that is, $\mb{w}_N(k+1) = \tilde{\mb{w}}\bigl( \mb{w}(k) \bigr)$ converges from $\mb{w}(0)$ to $\mb{w}_f$ in $N$ steps), the set unions in Proposition~\ref{prop:rec_feas} are finitely determined:
  \begin{equation*}
    \bigcup_{k=0}^{\infty} \mc{X}_N\bigl(\mb{w}(k)\bigr) = \bigcup_{k=0}^{N} \mc{X}_N\bigl(\mb{w}(k)\bigr)
  \end{equation*}
  \end{remark}

  Recursive feasibility leads to the following result.

\begin{proposition}[Exponential stability of $x_f(w_f)$]\label{prop:exp}
  Let $\mb{w}(0) \in \mc{W}$ and $x(0) \in \mc{X}_N(\mb{w}(0))$. If
  the disturbance sequence is updated as
  $\mb{w}(k+1) = \tilde{\mb{w}}\bigl(\mb{w}(k)\bigr)$, and the
  disturbance applied to the system is
  $w(k) = w\bigl(0;\mb{w}(k)\bigr)$, then the point $x_f(\mb{w}_f)$ is
  an exponentially stable equilibrium for the system
  $x^+ = Ax + B\kappa_N(x;\mb{w}) + w$.  The region of attraction is
  $\mc{X}_N(\mb{w}(0))$.
\end{proposition}

  Proposition~\ref{prop:exp} highlights the advantage of
  including the disturbance sequence in the prediction model and
  adjusting the terminal conditions accordingly: stability of a point
  is achieved. On the other hand, when the disturbance sequence is
  \emph{not} included in the prediction model, we have the following
  well-known result from the inherent robustness of nominal
  MPC~\cite{LAR+09,RM_mpc_book}, specialized to this setting.

  \begin{lemma}[Robust stability of nominal MPC]
    \label{lem:robstab}
    Let $\mb{w}(0) \in \mc{W}$ and
    $x(0) \in \Omega_\beta(\mb{w}=\mb{0}) \triangleq \bigl\{x :
    V^0_N(x;\mb{w}=\mb{0}) \leq \beta \bigr\} \subset
    \mc{X}_N(\mb{w}=\mb{0})$. If $\mc{W}$ is sufficiently small, then
    $\Omega_\beta(\mb{0})$ is a robust positively invariant set for
    $x^+ = Ax + B\kappa_N(x;\mb{0}) + w$. Moreover, the system states
    converge, in finite time, to a robust positively invariant set
    $\Omega_{\alpha}(\mb{0}) \subseteq \Omega_\beta(\mb{0})$, the size of
    which depends on $\mbb{W}$.
\end{lemma}
We do not attempt to qualify what is sufficiently small here; details
can be found in, \eg \cite[Chapter 3]{RM_mpc_book}. The main point is
that the states of the closed-loop system can be guaranteed to
converge to a neighbourhood of the origin, rather than an equilibrium
point, when the disturbance is omitted from the predictions.

\subsection{Persistently changing disturbance predictions}
\label{sec:main2}
We next consider the case where the disturbance sequence is permitted
to change over sampling instances, but where the rate of change is
limited. As a prerequisite to our developments, we consider the
composite state $z \triangleq (x,\mb{w})$ and define
\begin{equation*}
  \mc{Z}_N \triangleq \bigl\{ z = (x,\mb{w}) : \mb{w} \in \mc{W}, x \in \mc{X}_N(\mb{w}) \bigr\} \subset \mbb{R}^{n+nN}.
\end{equation*}
In view of the properties of $\mc{X}_N$ and $\mc{W}$, this is a
PC-set. The following lemma is a well established consequence of
linearity of the system, continuity of the cost, and compactness of
the constraint sets~\cite{LAR+09}.

\begin{lemma}[$\mc{K}$-continuity of the value function]
  The value function $V^0_N(\cdot)$ satisfies $\bigl|V^0_N(z_1) - V^0_N(z_2)\bigr|\leq\sigma_V(|z_1-z_2|)$ over
  $\mc{Z}_N$, with $\sigma_V$ a $\mc{K}$-function.
\end{lemma}
We exploit this continuity in bounding the rate of change of
disturbance sequences, as follows, and ultimately in establishing
robust stability.
\begin{assumption}\label{assump:w_rate}
  The disturbance sequence evolves as
  $\mb{w}^+ = \tilde{\mb{w}}(\mb{w}) + \Delta \mb{w}$, where
  $\Delta \mb{w} = \mb{w}^+ - \tilde{\mb{w}}(\mb{w}) \in \Delta \mc{W}
  \subseteq \mc{W}$. Moreover, $\Delta\mc{W}$ is chosen such that
  \begin{equation*}
    \lambda \triangleq \max \bigl\{ | \mb{w} - \mb{w}' | : \mb{w} \in \mc{W}, \mb{w}' \in \mc{W}, (\mb{w} - \mb{w}') \in \Delta\mc{W} \bigr\}
  \end{equation*}
  satisfies
  \begin{equation*}
    \lambda \leq \sigma_V^{-1}\bigl ( (\rho - \gamma) \alpha \bigr )
  \end{equation*}
  for $\rho \in (\gamma,1)$, where $\gamma = (1 - c_1/c_3) \in (0,1)$
  and $c_3 \geq c_2 \geq c_1 > 0$, and $\alpha > 0$ such that
  $\Omega^z_\alpha \triangleq \{z=(x,\mb{w}) : V_N^0(z) \leq \alpha\} \subset
  \mc{Z}_N$.
\end{assumption}

This assumption aims to bound the persistent change in the disturbance
predictions across sampling times, in order that the unanticipated
change is not so large that the stability of the controller is
lost. The constants $c_1$ and $c_2$ in this assumption are the same
ones from Assumption~\ref{assump:costs}, which bound the stage and
terminal costs. The constant $c_3$ is from the upper bound on the MPC
cost, derived in the proof of Proposition~\ref{prop:exp}; $\gamma$ is
the rate of the value function descent,
$V^0_N(x^+,\tilde{\mb{w}}(\mb{w})) \leq \gamma V^0_N(x;\mb{w})$, in
this result.

We then have the following, main result.
\begin{theorem}\label{thm:main}
  Suppose Assumptions~\ref{assump:contr}--\ref{assump:dist},
  \ref{assump:terminal_law}--\ref{assump:w_rate} hold, and let
  $z(0) = \bigl(x(0),\mb{w}(0)\bigr) \in \Omega^z_\beta = \{(x,\mb{w}) : V^0_N(z)
  \leq \beta\} \subset \mc{Z}_N$, for some $\beta \geq \alpha$. The set $\Omega^z_\beta$
  is positively invariant for the composite system
  \begin{align*}
    x^+ &= Ax+B\kappa_N(x;\mb{w}) + w, \\
    \mb{w}^+ &\in \tilde{\mb{w}}(\mb{w}) \oplus \Delta \mc{W}
\end{align*}
and the states of the system enter $\Omega^z_\alpha$ in finite time and
remain therein.
\end{theorem}
\begin{corollary}
  If
  $x(0) \in \Omega_\beta\bigl(\mb{w}(0)\bigr) = \{ x : \bigl(x,\mb{w}(0)\bigr) \in \Omega^z_\beta\} \subset
  \mc{X}_N\bigl(\mb{w}(0))$, then the system state $x(k)$ remains
  within
  $\Omega_\beta\bigl(\mb{w}(k)\bigr)= \{ x : \bigl(x,\mb{w}(k)\bigr) \in \Omega^z_\beta\} \subset
  \mc{X}_N\bigl(\mb{w}(k)\bigr) $ provided that the disturbance
  sequence update rate is limited as specified. Morever, all
  optimization problems remain feasible, and for all $k$ after some finite time $k'$, the state enters and remains in a set $\Omega_\alpha\bigl(\mb{w}(k')\bigr) =\{ x : \bigl(x,\mb{w}(k')\bigr) \in \Omega^z_\alpha\}$ where $\alpha \leq \beta$.
\label{thm:feas}
\end{corollary}

\begin{corollary}
  The region of attraction for the controlled system is
  \begin{equation*}
    \bigcup_{k=0}^{\infty} \Omega_\beta\bigl(\mb{w}(k)\bigr) \ \text{with} \ \mb{w}(k+1) \in \tilde{\mb{w}}\bigl(\mb{w}(k)\bigr) \oplus \Delta \mc{W}. 
    \end{equation*}
  \end{corollary}

  Given $\mb{w}(0)$, the set $\Omega_\beta\bigl(\mb{w}(0)\bigr)$ can
  be chosen to be the largest sublevel set of the value function
  within $\mc{X}_N\bigl(\mb{w}(0)\bigr)$; our result says then that
  the system converges to the smallest sublevel set that satisfies
  Assumption~\ref{assump:w_rate}. Comparing this result with the
  result of Lemma~\ref{lem:robstab}, the potential advantage of the
  disturbance sequence inclusion becomes clear: the region of
  attractions with and without the disturbance predictions are---using
  the same terminal sets in each optimal control
  problem---$\bigcup_{k=0} \Omega_\beta\bigl(\mb{w}(k)\bigr)$ and
  $\Omega_\beta(\mb{0})$ respectively. Care must be taken, however,
  before conclusions are made, because a conventional nominal
  formulation (omitting the disturbance) permits the use of a larger
  terminal set ($\mbb{X}_f \subseteq \mbb{X}$ compared with
  $\bar{\mbb{X}}_f \subseteq \beta_x \mbb{X}$). The example in the next
  section shows that, despite this, the proposed scheme can lead to a
  larger robust region of attraction.

  \begin{remark}[Comparison with robust MPC]
    We have studied a nominal formulation including disturbance
    predictions and analysed its inherent robustness to persistently
    changing predictions. Theorem~\ref{thm:main}, based on
    Assumption~\ref{assump:w_rate}, offers interesting theoretical
    insight but is not constructive and, thus, is of limited practical
    use: several constants and sublevel sets of the value function
    would need to be determined, which has to be done
    numerically. Robust tracking MPC techniques, which could also be
    applied to this problem, offer an attractive alternative, with
    robust-by-design guarantees and a region of feasibility that is
    equal to the region of attraction and possible to characterize
    explicitly. However, robust techniques rely on constraint
    restrictions and robust invariant sets~\cite{Mayne14}, neither of
    which a nominal formulation requires, which introduces a degree of
    complexity to the design.
    \end{remark}

\section{Illustrative examples}
\label{sec:examples}
Consider the neutrally stable dynamics
\begin{equation}
  x^+ = \begin{bmatrix} 1 & 1\\ 0 & 1\end{bmatrix} x + \begin{bmatrix} 0.5\\ 1\end{bmatrix}u +  w
\label{eq:example_dynamics}	
\end{equation}
with state and input constraints
$\mbb{X} = \{x\in\Rset^2 : \norm{x}_\infty\leq 10\}$,
$\mbb{U} = \{u\in\Rset : |u|\leq 3\}$. The disturbance takes values in
the C-set
$\mbb{W} = \{w= (w_1,w_2)\in\Rset^2\colon|w_1|\leq 2,\ w_1=w_2\}$. The
nominal stage and terminal costs are quadratic:
\begin{align*}
	\bar{\ell}(x,u) = x^\top Q x + u^\top S u && \bar{V}_f(x) = x^\top P x,
\end{align*}
where $Q = I$, $S = 1$, and $P$ is solution of the Lyapunov equation
$(A+BK_f)^\top P (A+BK_f) - P = -(Q + K_f^\top S K_f)$ for a
stabilizing gain $K_f\in\Rset^{m\times n}$ characterizing the nominal
terminal control law $\bar{\kappa}_f(x) = K_f x$. The matrices $\Phi$
and $\Pi$ are determined according to \eqref{eq:Phi_Pi}, respectively, using the nominal terminal control law,
$\Phi = A_{K_f} = A+BK_f$ and $\Psi = (I-A_{K_f})^{-1}$. The nominal
terminal set, $\bar{\mbb{X}}_f$, and scaling constants $\beta_x$,
$\beta_u$, $\alpha_x$ and $\alpha_u$ are designed in line with the
requirements of Assumptions~\ref{assump:terminal_law} and
\ref{assump:alpha}: $\bar{\mbb{X}}_f$ is the maximal constraint
admissible invariant set for $x^+ = A_{K_f} x$ within the state set
$\beta_x \mbb{X}$ and input set $\beta_u \mbb{U}$, while
$\Psi \mbb{W}_f$ and $K_f\Psi\mbb{W}_f$ must fit within
$\alpha_x \mbb{X}$ and $\alpha_u \mbb{U}$ respectively. With the
assumption that $\mbb{W}_f=\mbb{W}$, we obtain $\beta_x = 0.672$,
$\beta_u = 0.7$, $\alpha_x = 0.328$ and $\alpha_u = 0.3$. 

We consider controlling the system from different initial states using
a horizon of $N=3$ and the following disturbance predictions at each
sampling instant $k$:
\begin{equation*}
  \mb{w}(k) = 	\begin{cases} 
    \mb{w}_0 = \{+0.9,-0.9,-0.9\} & \mod (k,5)=0 \\
    \mb{w}_1 = \{+0.9,+0.9,+0.9\} & \mod (k,5)=1 \\
    \mb{w}_2 = \{-0.9,-0.9,-0.9\} & \mod (k,5)=2\\ 
    \mb{w}_3 = \{-0.9,+0.9,-0.9\} & \mod (k,5)=3\\
    \mb{w}_4 = \{0,0,0\} & \mod (k,5)=4\\
  \end{cases}
\end{equation*}
The disturbance prediction cycles persistently between five different
sequences, while the disturbance applied to the plant at time $k$ is
always the first in the sequence $\mb{w}(k)$: \ie
$w(k) = w\bigl(0;\mb{w}(k)\bigr)$, so that ${w}(0) = +0.9$,
$w(1) = +0.9$, $w(2) = -0.9$, \emph{etc}. Following
Assumption~\ref{assump:w_rate}, the value of the maximum disturbance
variation is $\lambda = 3.17$.

\begin{figure}
\centering\footnotesize
\input{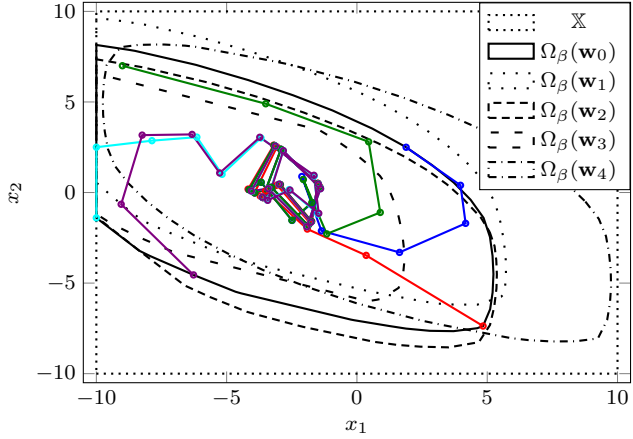}
\caption{Closed-loop trajectories of the system under switching
  disturbances, from different initial states, and the level sets $\Omega_\beta(\mb{w}_i), i=0\dots 4$, for $R=100$, corresponding to the disturbance sequences $\mb{w}_i, i=0\dots 4$.}
\label{fig:trajectories_w}
\end{figure}

\begin{figure}
\centering\footnotesize
\definecolor{cyan}{rgb}{0.00000,1.00000,1.00000}
\begin{tikzpicture}

\begin{axis}[%
enlargelimits=false,
width=1\linewidth,
height=0.75\linewidth,
xmin=-10.5,
xmax=10.5,
xlabel={$x^1_i$},
ymin=-10.5,
ymax=10.5,
ylabel={$x^2_i$},
legend style={at={(1,1)},anchor=north east},
]
\addplot[area legend,dotted,thick,draw=black,]
table[row sep=crcr] {%
x	y	z\\
10	-10 0\\
-10	-10 0\\
-10 10 0\\
10	10 0\\
}--cycle; 
\addlegendentry{$\mbb{X}$}
\addplot[area legend,solid,thick,draw=black,]
table[row sep=crcr] {%
9.99999999999996	-0.770464114370073	0\\
10	-4.80988356686672	0\\
9.99999999999996	-9.92636116614451	0\\
9.99551756686013	-10	0\\
8.31340161373562	-10	0\\
-3.84929732569015	-8.99999999999998	0\\
-4.1759785582847	-8.91201072085763	0\\
-4.95957428440695	-8.5202128577965	0\\
-5.10723878372673	-8.39276121627327	0\\
-9.36492930191062	-4.13507069808938	0\\
-10	-3.48786087719138	0\\
-10	-3.32982337728821	0\\
-10	3.62381995185792	0\\
-10	9.80897356413718	0\\
-9.99999999999995	9.92636116639544	0\\
-9.9955175663868	10	0\\
-6.94566017635258	10	0\\
3.84929732362986	8.99999999999999	0\\
4.17597855854603	8.91201072072698	0\\
4.95957428468711	8.52021285765642	0\\
5.1072387836683	8.3927612163317	0\\
9.36492930159041	4.13507069840959	0\\
9.98793579435088	3.50000000000001	0\\
}--cycle; \label{fig:XN}
\addlegendentry{$\bigcup_{\mb{w}}\Omega_\beta(\mb{w})$}

\addplot[area legend, dash pattern=on 2pt off 5pt on 2pt off 4pt, thick, dashdotted, draw=black, fill = none, fill opacity = 0.2]
table[row sep=crcr] {%
8.76638629678177	-0.499999999999995	0\\
9.28246320144738	-1.45560853070427	0\\
9.55768031860404	-2.41121706140854	0\\
9.67625320715062	-3.36682559211281	0\\
9.73922823004052	-4.28470600371499	0\\
9.75153058442822	-5.20258641531717	0\\
9.70843145096914	-6.12046682691934	0\\
9.60188071393684	-7.03834723852152	0\\
9.25197057427022	-8.04314621266324	0\\
9.19285567733442	-8.06711655490507	0\\
8.38571135466885	-8.21194299920045	0\\
7.57856703200327	-8.22062957498801	0\\
7.39379491626107	-8.21615378830531	0\\
6.58564364251693	-8.14845412800547	0\\
5.93952301420556	-8.04314621266324	0\\
5.77749236877279	-8.00164090812505	0\\
3.35303854754038	-7.33302699789362	0\\
2.54488727379624	-7.0901673025514	0\\
2.39882983108389	-7.03834723852152	0\\
0.00342082180854227	-6.12046682691934	0\\
-1.73673600005214	-5.30014936981578	0\\
-1.93614207711717	-5.20258641531717	0\\
-3.5934003552656	-4.28470600371499	0\\
-4.95553579264413	-3.36682559211281	0\\
-6.19831049139542	-2.41121706140854	0\\
-7.20906606861061	-1.45560853070427	0\\
-8.04756052194315	-0.499999999999995	0\\
-8.76638629664519	0.499999999999996	0\\
-9.28246320106036	1.45560853070427	0\\
-9.55768031860404	2.41121706140854	0\\
-9.67625320715062	3.36682559211282	0\\
-9.73922823004052	4.284706003715	0\\
-9.75153058442821	5.20258641531718	0\\
-9.70843145096914	6.12046682691936	0\\
-9.60188071393962	7.03834723852154	0\\
-9.1928556773344	7.83608642986002	0\\
-8.95990121016252	8.04314621266324	0\\
-8.38571135466881	8.15067229888078	0\\
-7.57856703200321	8.17338142366037	0\\
-7.39379491626107	8.17019974113046	0\\
-6.58564364251694	8.10741623526078	0\\
-6.19131252899538	8.04314621266324	0\\
-5.77749236877281	7.93714322998581	0\\
-3.35303854754041	7.2941526318925	0\\
-2.54488727379628	7.05486996209068	0\\
-2.49831715088891	7.03834723852154	0\\
-0.108740318102349	6.12046682691936	0\\
1.7367360000521	5.25824609938064	0\\
1.85049720492214	5.20258641531718	0\\
3.50997846892601	4.284706003715	0\\
4.95553579293637	3.36682559211282	0\\
6.19831049147648	2.41121706140854	0\\
7.20906606857874	1.45560853070427	0\\
8.04756052199178	0.499999999999996	0\\
}--cycle; \label{fig:ON_R_s}
\addlegendentry{$\Omega^\ts{standard}_\beta$}
\end{axis}
\end{tikzpicture}%
\caption{The union of level sets $\Omega_\beta(\mb{w})$ over $\mb{w}\in \mc{W}$ for $\beta=100$, and the corresponding level set $\Omega_\beta^\ts{standard}$ for a conventional MPC formulation.}
\label{fig:feas_region}
\end{figure}
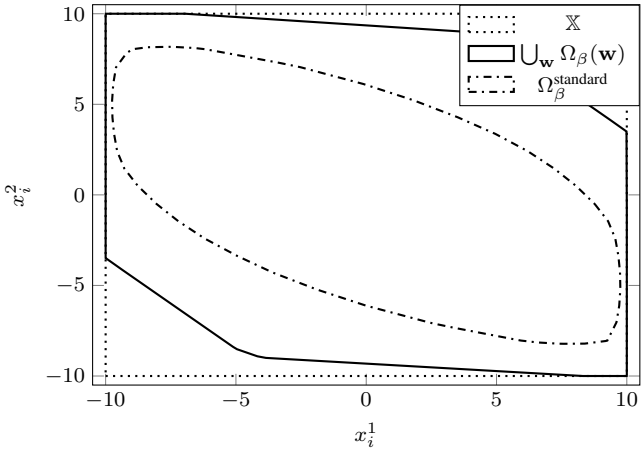

Figure~\ref{fig:trajectories_w} shows the trajectories of the
closed-loop system from five different initial states, together with
the level sets of the value function,
$\Omega_\beta(\mb{w}_i) = \{ x : V^0_N(x;\mb{w}_i) \leq \beta \}$ for
$i=0\dots 4$ and $\beta=100$. Each of the trajectories begins in
$\Omega_\beta(\mb{w}_0)$---the level set corresponding to the
disturbance sequence at $k=0$---and each subsequent state lies within
the appropriate $\Omega(\mb{w}(k))$; however,
$x(k) \in \Omega_\beta(\mb{w}(k))$ does not imply
$x(k+1) \in \Omega_\beta(\mb{w}(k))$, as can be seen with the
trajectory initialized at $x(0) = [1.9 \ 2.5]^\top$, which begins in
$\Omega_\beta(\mb{w}_0)$ and moves outside to
$\Omega_\beta(\mb{w}_1)$. Since the disturbances and their predictions
are switching between non-decaying values, the states do not converge
to zero, but rather to a neighbourhood of the origin.

Figure~\ref{fig:feas_region} compares the union of $\beta=100$
sublevel sets over $\mb{w} \in \mc{W}$ with the corresponding level
set for a conventional MPC controller, omitting the disturbance. As we
pointed out in the previous section, the region of attraction for the
conventional controller is at least as large as
$\Omega_\beta(\mb{w}=\mb{0})$ in the proposed approach, because the
latter requires the terminal set to fit within $\beta_x \mbb{X}$,
rather than merely $\mbb{X}$ as the conventional controller
requires. Despite this, the \emph{overall} region of attraction for
the proposed approach---as the union of $\Omega_\beta(\mb{w})$,
following Corollary~\ref{thm:feas}---is larger.
\section{Conclusions}
\label{sec:conc}

We have studied the use of available disturbance predictions within a
conventional nonlinear MPC formulation for regulation. Modifications
to standard terminal conditions were presented. For unchanging
disturbance predictions, recursive feasibility is guaranteed and
exponential stability of the closed-loop system is established around
an equilibrium point close to the origin. For arbitrarily changing
disturbance sequences, stability of a robust positively invariant set
is guaranteed, the size of which is related to the permitted
step-to-step change of the disturbance sequence.


\appendix
\subsection{Proof of Proposition~\ref{thm:terminal_set_inv}}
\label{sec:proof1}

Consider some $w_f \in \mbb{W}_f$. The point
$x_f = x_f(w_f) = \Psi  w_f = (I - \Phi)^{-1} w_f$ exists and is unique
by Lemma~\ref{lem:xeue}. Let $\mbb{X}_f(w_f) = \bar{\mbb{X}}_f \oplus \{x_f\}$, and consider
some $x \in \mbb{X}_f(w_f)$ and a corresponding
$z = x - x_f \in \bar{\mbb{X}}_f$. The successor states are
$z^+ = Az + B\bar{\kappa}_f(z)$, which is in $\bar{\mbb{X}}_f$ by
construction, and $x^+ = Ax + B\kappa_f(x;w_f) + w_f$, which, using
the control law definition
$\kappa(x;w_f) = \bar{\kappa}_f(x-x_f) + \Pi x_f$, may be
rewritten as
\begin{equation*}
 \begin{split}
   x^+ &= Ax + B\bar{\kappa}_f(x-x_f) + B\Pi x_f + w_f\\
   & = A(z+x_f) + B\bar{\kappa}_f(z) + B\Pi  x_f + w_f\\
   & = Az + B\bar{\kappa}_f(z) + (A + B\Pi)x_f + w_f\\
   & = z^+ + \Phi x_f + w_f\\
   & = z^+ + x_f
 \end{split}
\end{equation*}
where the last line follows from
$\Phi x_f + w_f = \Phi\Psi w_f + w_f = (\Phi\Psi + I)w_f$, and since
$\Psi = (I - \Phi)^{-1} = (I-\Phi)^{-1} \Phi + I$, then
$(\Phi\Psi + I) = \Psi$ and $\Psi w_f = x_f$. Then
$x^+ \in \mbb{X}_f(w_f)$ because $z^+ \in \bar{\mbb{X}}_f$. This
establishes positive invariance of $\mbb{X}_f(w_f)$, for any
fixed $w_f \in \mbb{W}_f$, under $u = \kappa_f(x;w_f)$.

To prove constraint admissibility, by Assumption~\ref{assump:alpha},
we have that $\Psi \mbb{W}_f \subseteq \alpha_x\mbb{X}$ and
$\Pi \Psi \mbb{W}_f \subseteq \alpha_u \mbb{U}$, with
$\alpha_x, \alpha_u \in [0,1)$. On the other hand, by
Assumption~\ref{assump:terminal_law},
$\bar{\mbb{X}}_f \subseteq \beta_x \mbb{X}$ and
$\bar{\kappa}_f(\bar{\mbb{X}}_f) \subseteq \beta_u \mbb{U}$, with
$\beta_x, \beta_u \in [0,1)$. Then
$\mbb{X}_f(\mbb{W}_f) = \bar{\mbb{X}}_f \oplus \Psi\mbb{W}_f \subseteq
\beta_x \mbb{X} \oplus \alpha_x \mbb{X} \subseteq \mbb{X}$, if
$\alpha_x + \beta_x \leq 1$. Similarly,
$\kappa_f(\mbb{X}_f;\mbb{W}_f) = \bar{\kappa}_f(\bar{\mbb{X}}_f)
\oplus \Pi \Psi \mbb{W}_f \subseteq \beta_u \mbb{U} \oplus \alpha_u
\mbb{U} \subseteq \mbb{U}$ if $\alpha_u + \beta_u \leq 1$.

Next, we prove the claimed properties of the functions
$\ell(\cdot,\cdot)$ and $V_f(\cdot)$. For some $w_f\in \mbb{W}_f$,
$x\in\mbb{X}_f(w_f)$ and corresponding
$z = x - x_f \in \bar{\mbb{X}}_f$, we have $\bar{V}_f(z^+) - \bar{V}_f(z) \leq -\bar{\ell}\bigl(z,\bar{\kappa}_f(z)\bigr)$ by construction (Assumption~\ref{assump:terminal_law}). Then, for $x^+ = Ax + B\kappa_f(x;w_f) + w_f$,
\begin{equation*}
 \begin{split}
   {V}_f(x^+) - {V}_f(x) &= \bar{V}_f(x^+ - x_f) - \bar{V}_f(x - x_f)\\
   &= \bar{V}_f(z^+ + x_f - x_f) - \bar{V}_f(z + x_f - x_f)\\
   & = \bar{V}_f(z^+) - \bar{V}_f(z) \\
   & \leq -\bar{\ell}\bigl(z,\bar{\kappa}_f(z)\bigr)\\
   &\quad = - \bar{\ell}\bigl( x-x_f, \bar{\kappa}_f(x - x_f) \bigr)\\
   &\quad = - \bar{\ell}\bigl( x-x_f, {\kappa}_f(x;w_f) - \Pi x_f \bigr)\\
   &\quad = - \ell(x,u; w_f) \ \text{where} \ u = \kappa_f(x,w_f)
 \end{split}
\end{equation*}
as required. Finally, the asymptotic stability of $x_f$ for the system
$x^+ = Ax + B\kappa_f(x;w_f) + w_f$ follows from the bounds on
$V_f(\cdot)$: for all $x \in \mbb{X}_f({w}_f)$
\begin{gather*}
 c_1 | x - x_f |^a  \leq V_f(x;w_f) \leq c_2 |x - x_f|^a \\
 {V}_f(x^+;w_f) - {V}_f(x;w_f) \leq - c_1 | x - x_f |^a.
 \end{gather*}
 Thus, $x \to x_f$ in the limit. \IEEEQED

\subsection{Proof of Proposition~\ref{prop:rec_feas}}
\label{sec:proof2}

For (i), given $x \in \mc{X}_N(\mb{w})$ there exists a
$\mb{u}(x;\mb{w}) \in \mc{U}(x;\mb{w})$ with associated state
predictions
$\mb{x}(x;\mb{w}) = \bigl\{x^0(0;x,\mb{w}), x^0(1;x,\mb{w}), \dots,
x^0(N;x,\mb{w})\bigr\}$ with $x^0(0;x,\mb{w}) = x$. The successor
state $x^+ = Ax + Bu^0(0;x,\mb{w}) + w(0;\mb{w}) = x^0(1;x,\mb{w})$,
and so, by Proposition~\ref{thm:terminal_set_inv}, the sequences
\begin{align*}
 \begin{split}
   \tilde{\mb{x}}(x^+;\mb{w}) &= \bigl\{x^0(1;x,\mb{w}), \dots, x^0(N;x,\mb{w}), \\ &\quad Ax^0(N;x,\mb{w}) + B\kappa_f\big(x^0(N;x,\mb{w});w_f\bigr) +  w_f \bigr\}\end{split}\\
 \begin{split}
 \tilde{\mb{u}}(x^+;\mb{w}) &= \bigl\{u^0(1;x,\mb{w}), \dots, u^0(N;x,\mb{w}), \\&\quad \kappa_f\big(x^0(N;x,\mb{w});w_f\bigr) \bigr\}\end{split}\\
 \tilde{\mb{w}}(\mb{w}) &= \bigl\{ w(1), w(2), \dots, w(N-2), w(N-1) \bigr\} 
 \end{align*}
 are feasible for all constraints that define $\mc{U}_N\bigl(x^+,\tilde{\mb{w}}(\mb{w})\bigr)$; in fact, the same solution omitting the terminal steps is also feasible for all constraints that define $\mc{U}_{N-1}\bigl(x^+,\tilde{\mb{w}}(\mb{w})\bigr)$.  Thus, $x^+ \in \mc{X}_{N-1}(\mb{w}^+) \subseteq \mc{X}_{N}(\mb{w}^+)$ if $\mb{w}^+ = \tilde{\mb{w}}(\mb{w})$.

Part (ii) follows directly from the above and recursion. If $x \in \mc{X}_N(\mb{w})$ and $x^+ \in \mc{X}_{N-1}(\mb{w}^+) \subseteq  \mc{X}_N(\mb{w}^+)$ under $\mb{w}^+ = \tilde{\mb{w}}(\mb{w})$ and $u = \kappa_N(x;\mb{w})$, then, given $x(0) \in \mc{X}_N\bigl(\mb{w}(0)\bigr)$ and the disturbance update rule $\mb{w}(k+1) = \tilde{\mb{w}}\bigl(\mb{w}(k)\bigr)$, the state trajectory $\bigl\{ x(k) \big\}_k$ remains within the union of $\mc{X}_N\bigl( \mb{w}(k) \bigr)$. Hence, the latter set is positively invariant for $x^+ = Ax + B\kappa_N(x;\mb{w}) +  w$, and control invariant for $x^+ = Ax + Bu +  w$ and $\mbb{U}$.

The nested property of the set union follows from the key observation
that, under the tail-updating law, $\mb{w}_N(k+1) = \mb{w}_{N-1}(k)$. Thus, since $x(k) \in \mc{X}_N\bigl(\mb{w}_N(k)\bigr)$ implies $x(k+1) \in \mc{X}_{N-1}\bigl(\mb{w}_{N-1}(k)\bigr) \subseteq \bigcup_{k} \mc{X}_{N-1}\bigl(\mb{w}_{N-1}(k)\bigr) $, it also implies $x(k+1) \in \mc{X}_N\bigl(\mb{w}_N(k+1)\bigr) \subseteq \bigcup_{k} \mc{X}_N\bigl(\mb{w}_N(k)\bigr)$. Therefore, $\bigcup_{k} \mc{X}_N\bigl(\mb{w}_N(k)\bigr) \supseteq \bigcup_{k} \mc{X}_{N-1}\bigl(\mb{w}_{N-1}(k)\bigr)$. \IEEEQED

\subsection{Proof of Proposition~\ref{prop:exp}}
\label{sec:proof3}

From the definitions of $V_f$, $\ell$ and $V_N$, and the bounds in
Assumption~\ref{assump:costs}, we have, for all $x \in \mc{X}_N(\mb{w})$
\begin{equation*}
 V^0_N(x;\mb{w}) \geq \ell\bigl(x,\kappa_f(x;w_f); w_f\bigr) \geq c_1 |x - x_f|^a,
\end{equation*}
while the fact that the costs are continuous and the sets $\mbb{X}_f$,
$\mbb{X}$, $\mbb{U}$ are PC-sets means there exists a $c_3 \geq c_2 > 0$ such
that
\begin{multline*}
 V^0_N(x;\mb{w}) \leq V_f(x;w_f) \leq c_2 |x - x_f|^a \ \text{for all} \ x \in \mbb{X}_f({w}_f) \\\implies
 V^0_N(x;\mb{w}) \leq c_3 |x - x_f|^a \ \text{for all} \ x \in \mbb{X}_N(\mb{w})
\end{multline*}

Recursive feasibility and the descent property of $V_f$ yields, for all $x \in \mbb{X}_N(\mb{w})$,
\begin{equation*} {V}^0_N(x^+;\tilde{\mb{w}}) \leq {V}^0_N(x;\mb{w}) -
 \ell\bigl(x,\kappa_f(x;w_f); w_f\bigr) \leq \gamma V^0_N(x;\mb{w})
\end{equation*}
where $\gamma \triangleq (1-c_1/c_3) \in (0,1)$ (since
$c_3 > c_1 > 0$). Thus, ${V}^0_N\bigl(x(k);\mb{w}(k)\bigr) \leq \gamma^k {V}^0_N\bigl(x(0);\mb{w}(0)\bigr)$, and, because $c_1 |x - x_f |^a \leq V_N^0(x;\mb{w}) \leq  c_3|x - x_f|^a$,
 \begin{multline*}
   c_1 |x(k) - x_f|^a \leq \gamma^k c_3 |x(0) - x_f|^a \implies \\|x(k) - x_f| \leq c \delta^k |x(0) - x_f|
 \end{multline*}
 where $c \triangleq (c_3/c_1)^{1/a} > 0$ and $\delta \triangleq \gamma^{1/a} \in (0,1)$.
\IEEEQED
\subsection{Proof of Theorem~\ref{thm:main}}

Consider some $z = (x,\mb{w}) \in \Omega^z_N$. We have $V^0_N\bigl(x^+,\tilde{\mb{w}}(\mb{w})\bigr) \leq \gamma V^0_N(x,\mb{w}) \leq \gamma \beta$. By $\mc{K}$-continuity,
\begin{equation*}
V^0_N(x^+,\mb{w}^+) \leq V^0_N\bigl(x^+,\tilde{\mb{w}}(\mb{w})\bigr) + \sigma_V \bigl(\bigl| \mb{w}^+ - \tilde{\mb{w}}(\mb{w}) \bigr|\bigr).
\end{equation*}
Therefore, since $ \sigma_V \bigl(\bigl| \mb{w}^+ - \tilde{\mb{w}}(\mb{w}) \bigr|\bigr) \leq \sigma_V (\lambda) \leq (\rho-\gamma) \alpha$,
\begin{equation*}
V^0_N(x^+,\mb{w}^+) \leq \gamma \beta + (\rho - \gamma) \alpha \leq \rho \beta < \beta.
\end{equation*}
Moreover, if $\bigl(x(0),\mb{w}(0)\bigr) \in \Omega^z_\beta \setminus \Omega^z_\alpha$, then
\begin{equation*}
  \begin{split}
    V^0_N\bigl(x(1),\mb{w}(1)\bigr) &\leq  \gamma V^0_N\bigl(x(0),\mb{w}(0)\bigr) + (\rho-\gamma) V^0_N\bigl(x(0),\mb{w}(0)\bigr)  \\
    &\leq   \rho V^0_N\bigl(x(0),\mb{w}(0)\bigr).
    \end{split}
  \end{equation*}
  Consequently, $V^0_N\bigl(x(k),\mb{w}(k)\bigr) \leq \rho^k \beta$, from
  which it follows that $V^0_N\bigl(x(k'),\mb{w}(k')\bigr) \leq \alpha$
  after some finite $k'$.  \IEEEQED

\bibliography{\myreferences}
%
 
\end{document}